\documentclass[preprint]{aastex6}
\turnoffedit
\begin{document}


\title{Polarized disk emission from Herbig Ae/Be stars observed using Gemini Planet Imager: HD 144432, HD 150193, HD 163296, and HD 169142}


\author{John D. Monnier\altaffilmark{1}, Tim J. Harries\altaffilmark{2},
Alicia Aarnio\altaffilmark{1}, 
Fred C. Adams\altaffilmark{1}, 
Sean Andrews\altaffilmark{3}, 
Nuria Calvet\altaffilmark{1},
Catherine Espaillat\altaffilmark{4}, 
Lee Hartmann\altaffilmark{1},
Sasha Hinkley\altaffilmark{2}, 
Stefan Kraus\altaffilmark{2},  
Melissa McClure\altaffilmark{5}, 
Rebecca Oppenheimer\altaffilmark{6},
Marshall Perrin\altaffilmark{7}, and
David Wilner\altaffilmark{3}
}
\altaffiltext{1}{Astronomy Department, University of Michigan, Ann Arbor, MI 48109, USA}
\altaffiltext{2}{University of Exeter, Exeter, United Kingdom}
\altaffiltext{3}{Harvard-Smithsonian Center for Astrophysics, Cambridge, MA 91023 USA}
\altaffiltext{4}{Boston University, Boston, MA}
\altaffiltext{5}{European Southern Observatory, Garching, Germany}
\altaffiltext{6}{American Museum of Natural History, New York, NY USA}
\altaffiltext{7}{Space Telescope Science Institute, Baltimore, MD USA}

\begin{abstract}
In order to look for signs of on-going planet formation in young disks, we carried out the first J-band polarized emission imaging of the Herbig Ae/Be stars HD~150193, HD~163296, and HD~169142 using the Gemini Planet Imager (GPI), along with new H band observations of HD~144432.
We confirm the complex ``double ring'' structure for the nearly face-on system HD~169142 first seen in H-band, finding the outer ring to be substantially redder than the inner one in polarized intensity.   Using radiative transfer modeling, we developed a physical model that explains the full spectral energy distribution (SED) and J- and H-band surface brightness profiles, suggesting that the differential color of the two rings \edit1{could come} from reddened starlight traversing the inner wall and \edit1{may not require} differences in grain properties. 
In addition, we clearly detect an elongated, off-center ring in HD~163296 (MWC~275), locating the scattering surface to be 18~AU above the midplane at a radial distance of 77~AU, co-spatial with a ring seen at 1.3mm by ALMA linked to the CO snow line.  Lastly,  we report a weak tentative detection of scattered light for HD~150193 (MWC~863) and a non-detection for HD~144432; the stellar companion known for each of these targets has likely disrupted the material in the outer disk of the primary star.   For HD~163296 and HD~169142, the prominent outer rings we detect could be evidence for giant planet formation in the outer disk or a manifestation of large-scale dust growth processes possibly related to snow-line chemistry.
\end{abstract}

\keywords{techniques: polarimetric --- protoplanetary disks --- stars: pre-main sequence --- infrared: planetary systems --- radiative transfer  }



\section{Introduction} 
\label{sec:intro}
While astronomers have detected thousands of exoplanets (2951 confirmed planets to date on exoplanets.org), we still lack a predictive theory of planet formation that can explain their distributions in mass, orbital characteristics, and dependence on host star properties.  Many ingredients for planet formation -- such as the streaming instability, dust growth, dead zones, gravitational instability, core accretion, planetary migration, and more -- have been identified \edit1{\citep[e.g.,][]{johansen2007, birnstiel2010,  turner2008, pollack1996, boss1997, tanaka2002}} but there is no consensus as to their relative importance as numerical models struggle to match the increasingly rich and diverse constraints with only simple inputs.  

Fortunately we have more than just the final distribution of exoplanetary systems to learn from.  Indeed we can directly observe key stages of the planet formation process through high angular resolution imaging in \edit1{the closest} star forming regions.  \edit1{For young stars located about $\sim$100\,pc away, diffraction-limited imaging by 8-m class infrared telescopes  can measure faint polarized scattered light with $\sim$5~AU resolution, while infrared interferometers can measure thermal emission from hot dust (T$>$1000K) within 1~AU.}  \edit1{Most recently, the mm-wave interferometer ALMA has been commissioned with the capabilities to eventually image both gas and dust emission with $<$1~AU-resolution for some objects.  For instance, ALMA has already detected a series of nested gaps in the HL~Tau system \citep{alma2015} in a mode with $\sim$3~AU imaging resolution.}  The combination of infrared imaging and mm-wave imaging shows how the large and small grains can be decoupled by disk vortices in some cases\citep[e.g., IRS Oph 48;][]{vandermarel2013} or be more co-spatial in spiral arm structures in others\citep[e.g. SAO 206462;][]{garufi2013,perez2014}. 

Some of the features seen in the outer disks can be explained by interactions with unseen giant planets within the disk \cite[e.g., SAO 206462;][]{bae2016}. \edit1{ For instance, large cavities in disks have been observed in some systems, e.g., PDS~70 \citep{hashimoto2012}, 2MASS J16042165-2130284 \citep{canovas2016},  HD 100546 \citep{currie2015}, and these cavities may have been sculpted by unseen planets. }  Directed by theory,  searches have so far not been able to confirm the presence of young exoplanets in most cases and some researchers look toward other explanations involving snow lines \citep{zhang2016} and/or dust evolution \citep{birnstiel2015}.  More generally, direct imaging surveys for exoplanets around a wider range of stars also are starting to provide constraints on the prevalence of giant planets. \edit1{\citet{brandt2014} and}  \citet{galicher2016} report  giant planets are rare beyond 20 AU (present for $<$1\% of systems); if true, then perhaps the rings and gaps  commonly seen in the outer regions of YSO disks are coming from something else (or possibly the young giant planets migrate inward as the disk evolves).

Within the next 10 years, astronomers will obtain dozens of high resolution images of young star disks in mm-wave emission (ALMA), scattered light (Subaru, GPI, SPHERE) and mid-infrared thermal emission (VLTI/MATISSE).  Here we present four new deep observations of well-known Herbig Ae/Be stars in polarized scattered light using the Gemini Planet Imager.  Our new data presently are the highest resolution and signal-to-noise images of complex features within 100 AU of these targets and we present a preliminary analysis and discuss our findings in the context of current debates on the giant planet formation in the outer solar system beyond 20~AU.

\section{Observations and Data Processing}
\label{data}
We report new imaging of Herbig Ae/Be stars using  the Gemini Planet Imager \citep[GPI;][]{macintosh2008,macintosh2014,poyneer2016} installed on Gemini South. In polarimetry mode \citep{perrin2015} with the adaptive optics system and an occulting spot, GPI can obtain high dynamic range imaging of scattered light from Y-K bands relying on the physics of scattering to deliver a  distinctive polarization pattern. \edit1{Light scattered from dust grains will be polarized with E-field vectors aligned perpendicular to the radial direction toward the star, while the light from the central star's PSF will be typically unpolarized or linearly polarized throughout the PSF. }

This paper collects data from two separate observing runs, one in 2014 (GS-2014A-SV-412) and one in 2015 (GS-2015A-Q-49).  For the data presented here, we utilized the standard GPI coronagraphic configurations (specifically 'J-coron' and 'H-coron'), including use of a coronographic spot (0.184'' diameter for J band and 0.246'' diameter for H band) and appropriate Lyot and apodizing pupil masks. We chose integration times to just avoid saturation of light around the spot, ranging from 15 to 30 seconds. We coadded either 2 or 4 frames together to accumulate 1~minute of on-source exposure time per file, a limit imposed by the rotating field-of-view in the GPI design.  We used the Wollaston prism mode and rotated the half-wave plate 22.5\arcdeg~between each 1~minute observation.  Table~\ref{table:targets} contains the information on the target stars while Table~\ref{table:obslog}  contains the Observing Log.  

\subsection{Polarization Analysis}
Here we outline the data reduction steps to extract polarimetric observables. The work was largely carried out using the IDL-based GPI pipeline version 1.4 along with custom routines written in IDL.  See \citet{perrin2014} and \citet{millar-blanchaer2016} for more detailed descriptions of the basic method outlined here.

Firstly, we searched the Gemini Data Archive for the best calibration files.  Specifically, we created darks from files taken close in time  and with the same integration time as used for the science data and the flats.  We used the standard pipeline recipe ``Calibration/Darks'' for this purpose.  Note that the darks and flats were not always taken on the same day as the science observations.   We then followed this step by using the daytime flat calibration files to calibrate spot locations using recipe ``Calibration/Calibrate Polarization Spots Locations - Parallel.'' We found that some of the defaults for this recipe changed between pipeline 1.3 and 1.4 which caused this step to fail originally, but the spot calibration was successful after returning to the defaults from v1.3. At this stage we also used the recipe ``Calibration/Create Low Spatial Frequency Polarized Flat-field'' for use in later steps.  Bad pixel maps and a few other calibration files were needed by the pipeline and these were downloaded from the Gemini GPI website.

Following these preliminaries, we proceed to reduce the actual science data.  All the individual exposures taken with the Wollaston Prism were processed using a slightly modified recipe ``PolarimetricScience/Simple Polarization Cube Extraction.''  
We removed the step that attempts to measure the flux using the satellite spots after concluding this procedure was not reliable enough for our use.  This recipe created a series of polarization datacube (``podc'') files.   At this stage, careful attention was paid that the center of the pattern was accurately measured by the pipeline using the primitive\footnote{"Primitive" is the term used by the GPI pipeline to refer to core analysis routines.} ``Measure Star Position from Polarimetry.''  Unfortunately this algorithm, based on the Radon Transform \citep[see description and tests in ][]{wang2014}, does fail when a companion is nearby (e.g., for HD~144432, HD~150193).  A custom and interactive version of this primitive was written to allow detailed masking of the companion and its diffraction features.  After testing the results of our routine against the standard primitive for single stars, we then re-analyzed all the polarization datacube  files using our own algorithm, modifying the header keywords PSFCENTX, PSFCENTY in FITS extension 1. We estimated an extra $\pm$0.5 pixel error on the centroid estimates when a companion was present. While we were not able to locate the position of HD~144432B since it was slightly off-chip, we can report a new astrometric position for HD~150193B relative to A: 1.12\arcsec$\pm$0.02\arcsec at PA 223\arcdeg$\pm$1\arcdeg \citep[compare to 1.10\arcsec$\pm$0.03\arcsec ~at PA 225.0\arcdeg$\pm$0.8\arcdeg ~reported by][]{fukugawa2010}.

Next, the individual files were grouped in chunks of 8 files each separated by a 22.5\arcdeg~rotation of the half-wave plate and then analyzed using a slightly modified version of the recipe ``PolarimetricScience/Basic Polarization Sequence (From polarization datacubes).'' In our version, we \edit1{excluded} the primitive ``Subtract Mean Stellar Polarization'' in order to implement our own version later in the process.  As an output of this step, the pipeline used a singular value decomposition (SVD) method \citep[e.g.,][]{perrin2015} to estimate the Stokes $I,Q,U,V$ components from the set of 8 observations. 
 
\edit1{As emphasized in the last paragraph, we did not use the built-in primitive to subtract the stellar polarization but instead implemented this procedure ourselves and we will now explain our method in detail.  Indeed, it is important to remove the stellar/instrumental polarization in order to see the scattered light from circumstellar emission. Even} a 1\% linearly polarized PSF will dominate over the polarized dust emission beyond 0.5'' or so.  We experimented with a few different methods for estimating the underlying stellar polarization before removing it. While other workers have chosen to use the signal either behind or just outside the coronagraphic spot \citep[see the interesting study by][]{millar-blanchaer2016}, we used the azimuthally-averaged Stokes $I,Q,U,V$ surface brightness profile to estimate residual linear polarization, extracting the median value of $f_Q=Q/I$, $f_U=U/I$, $f_V=V/I$ within 1.2'' (for the single H-band dataset of HD~144432 we used a region within 0.4" based on visual inspection of the results).  Note that using flux behind the occulting spot gives similar results but our procedure demonstrated less variance when applied to a large test dataset.  \deleted{We do not attempt to optimize this process by dithering the stellar position estimate as done by \cite{quanz2013} but fix the position based on the earlier Radon Transform estimate.} Once we have the $f_{Q,U,V}$ we can multiply this by the total intensity $I$ in each pixel to estimate the $Q,U,V$ contamination and subtract these contributions from the linear polarization.  For reference, we report the mean stellar linear polarization \edit1{($P_{\rm band} = (f^2_Q + f^2_U)^\frac{1}{2}$, $\Theta = \frac{1}{2}\arctan{\frac{U}{Q}}$)} that we removed (all angles are degrees East of North): 
HD~144432 $P_H=1.1\%$ at $\Theta=-4\arcdeg$,      
HD~150193 $P_J=3.0\%$ at $\Theta=56\arcdeg$, 
HD~163296 $P_J=0.8\%$ at $\Theta=31\arcdeg$, and
HD~169142 $P_J=0.5\%$ at $\Theta=-1.5\arcdeg$. 
For HD~163296 and HD~169142 the observed polarization angle varied $\sim10\arcdeg$ as a function of parallactic angle suggesting that these measurements are partially still contaminated by uncorrected instrumental effects and not totally intrinsic. \edit1{We refer the reader to the GPI instrumentation papers referenced in \S\ref{data} for more information on the systematic errors related to removal of the instrumental signature in the pipeline.}  That said, generally our values are broadly consistent with measurements at similar wavelengths: 
HD~144432 $P_I=1.8\%$ at $\Theta=20\arcdeg$ \citep{oudmaijer2001},
HD~150193 $P_I=4.6\%$ at $\Theta=60\arcdeg$ \citep{oudmaijer2001}, 
HD~163296 $P_I=0.2\%$ at $\Theta=45\arcdeg$ \citep{oudmaijer2001}, and
HD~169142 $P_I\sim0.3\%$ \citep{chavero2006}.  Also for comparison, \citet{hales2006} made the following report: HD~144432 $P_J=0.5\%$ at $\Theta=3\arcdeg$,      
HD~150193 $P_J=3.1\%$ at $\Theta=57\arcdeg$, 
HD~169142 $P_J=0.2\%$ at $\Theta=37\arcdeg$. 

Following subtraction of mean stellar polarization signal from the Stokes data cube (one for each group of 8 files), we then coadded multiple stokes datacubes spanning a range of parallactic angles. Lastly, we project the traditional Stokes $Q,U$ components (oriented relative to North/East) onto a radial basis set $Q_r,U_r$ based on the stellar position determined earlier in the processing.  In this procedure \citep[see derivation in][]{schmid2006,avenhaus2014,garufi2014,millar-blanchaer2016},  linear polarization vectors that are azimuthally-oriented around the center are positive in $Q_r$ space while radial vectors are negative.  Similarly polarization vectors oriented $\pm$45$\arcdeg$ from this are found in the $U_r$ component. This projection is very practical since single-scattering should be oriented around the stellar position and produce purely positive $Q_r$ signal, while noise can be both positive and negative. Furthermore,  miscalibrations (especially in the inner PSF halo) will produce residual $U_r$ signal that can be used to assess data quality and guard against false conclusions.  That said, we recognize the limitations of this presentation and refer to \citet{canovas2015} and  \citet{dong2016} for more sophisticated discussion of polarization signatures for optically-thick, more edge-on disks.

\begin{table}
\centering
\caption{Target List}
\label{table:targets}
\begin{tabular}{llllrrrrccc}
\hline
 &  & \colhead{RA}  & \colhead{Dec} & & \multicolumn{3}{c}{2MASS\tablenotemark{a}}   &\colhead{T$_{\rm eff}$} & \colhead{Meeus\tablenotemark{c}} & \colhead{Distance}   \\
\colhead{HD Number}& \colhead{Alias} & \colhead{J2000} & \colhead{J2000} & \colhead{R} & \colhead{J}  & \colhead{H} & \colhead{K$_s$} & \colhead{(K)\tablenotemark{b}} & \colhead{Group} & \colhead{(pc)} \\
\hline
HD~144432A &          & $16$ $06$ $57.95489$ & $-27$ $43$ $09.7880$ & 7.8 & 7.1 & 6.5 & 5.9 & 7500 & IIa & 145\tablenotemark{d}\\
HD~150193A & MWC~863A & $16$ $40$ $17.92287$ & $-23$ $53$ $45.1787$ & 8.4 & 6.9 & 6.2 & 5.5 & 9500 & IIa & 150\tablenotemark{e}\\
HD~163296  & MWC~275  & $17$ $56$ $21.28803$ & $-21$ $57$ $21.8700$ & 6.9 & 6.2 & 5.5 & 4.8 & 9200 & IIa & 119\tablenotemark{f}\\
HD~169142  & MWC~925  & $18$ $24$ $29.7787$  & $-29$ $46$ $49.371$  & 8.2 & 7.3 & 6.9 & 6.4 & 7500 & Ib  & 145\tablenotemark{g}  \\
\hline
\multicolumn{11}{l}{Refs. (a) \citet{2mass} (b) \citet{alecian2013} (c) \citet{meeus2001} (d) \citet{perez2004}  }\\ 
\multicolumn{11}{l}{(e) \citet{feigelson2003} (f) \citet{vanleeuwen2007} (g) \citet{sylvester1996} }
\end{tabular}
\end{table}

\begin{table}
\centering
\caption{Observing Log of Polarimetry Observations using Gemini Planet Imager. For these observations we used the default occulting spot, apodizer, and Lyot stop appropriate for the observing waveband. }
\label{table:obslog}
\begin{tabular}{llcccccc}
\hline
\colhead{UT Date} & \colhead{Target Name}   & \colhead{Filter} & \colhead{T$_{\rm int}$ (sec)}         & \colhead{N$_{\rm coadds}$} &  \colhead{N$_{\rm Frames}$\tablenotemark{a}} & \colhead{Airmass} & \colhead{Seeing (\arcsec)\tablenotemark{b}} \\
\hline
2014 April 23 & HD~150193A & J & 29.10 & 2 & 16 & 1.10$-$1.16 & 0.46$-$0.68 \\
2014 April 24 & HD~150193A & J & 29.10 & 2 & 16 & 1.09$-$1.16 & 0.65$-$0.86 \\
2014 April 24 & HD~163296  & J & 14.55 & 4 & 32 & 1.03$-$1.07 & 0.48$-$0.84 \\
2014 April 25 & HD~169142  & J & 29.10 & 2 & 64 & 1.00$-$1.40 & 0.58$-$1.11 \\
2014 April 25 & HD~91538   & J & 1.5 & 4 & 8 & 1.021$-$1.023 & 0.61$-$0.76 \\
2015 July 9 & HD~144432A & H & 29.10 & 2 & 24 & 1.11$-$1.21 & 0.89$-$1.40 \\
2015 July 9 & HR~6572    & H & 1.5 & 4 & 14 & 1.32$-$1.37 & 0.71$-$0.79 \\
\hline
\end{tabular}
\tablenotetext{a}{Here we refer to the number of frames used in the data reduction, where a frame consists of N$_{\rm coadds}$ images coadded with individual exposures times of T$_{\rm int}$ seconds at a single half-wave plate position.  A few recorded frames were unusable due to clouds or poor guiding behind the occulting spot. }
\tablenotetext{b}{Seeing column is value reported in headers based on DIMM measurements.}
\end{table}

\subsection{Flux Calibration}
Our data were taken during the early stages of Gemini Planet Imager commissioning and a full set of flux calibration data were not taken.  It has been difficult to independently verify some important throughput estimates needed to calibrate the surface brightness in physical units.  For J band, we used the PSF source HD~91538 (J mag 5.42) observed on 2014 Apr 25 with the Wollaston prism in J band but which had the adaptive optics loop off and had the coronagraph spot, apodizer, and Lyot stop out.  Laboratory tests demonstrated that we expect 2.69$\times$ less light through the system when inserting the apodizer and Lyot stop for J band mode \edit1{\citep[private communication Patrick Ingraham; additional information can be found in][]{maire2014}} and we applied this factor when calibrating our flux.  Unfortunately the companion stars for HD 144432 and HD~150193 were both either saturated or partially outside the field-of-view and could not be used for flux calibration.  For the 2015 H band data, we requested a PSF star observation to be taken with the same occulting spot, apodizer,  and Lyot stop as the polarization target and HR~6572 (H mag 5.74)  was observed with AO loop off.   Since these calibrations could be affected by cloud or atmospheric conditions \edit1{and the observations were taken under queue observing with no specific note about cirrus or possible thin clouds, we can expect at least 25\% error in our flux calibration. When our flux calibration is applied over many nights, the errors could be larger since atmospheric conditions vary more over a longer time base}.  A more robust flux calibration procedure will be adopted in future work since comparisons between J,H,K polarized surface brightness provide critical constraints in probing  dust size distributions.

\begin{figure}
\centering
\includegraphics[width=4in]{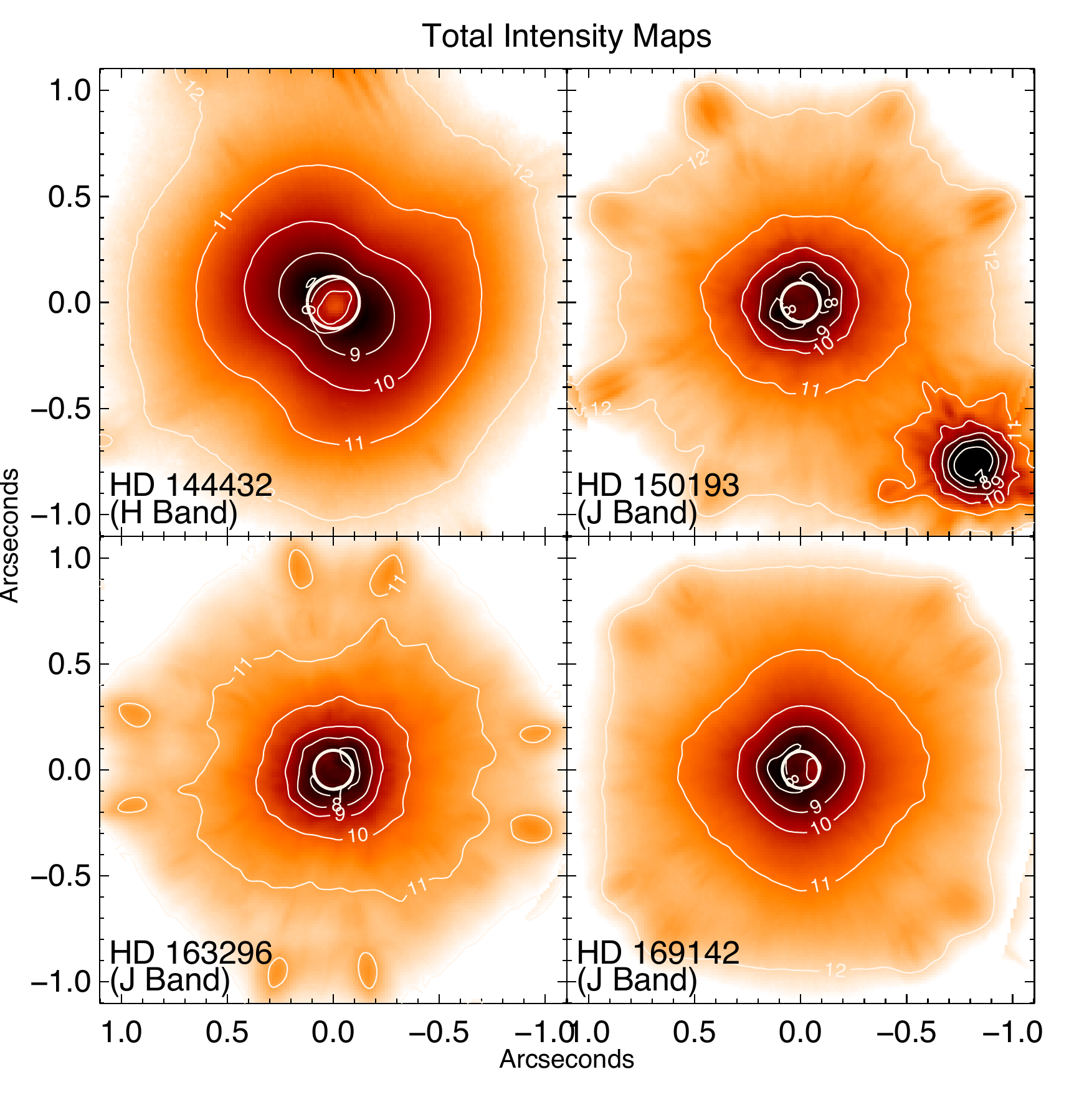}
\caption{Total intensity maps, measured by the Gemini Planet Imager, shown using a logarithmic color table for the four targets stars.   The intensity scales (maximum and minimum) are different for each panel and local surface brightness levels can be found as labeled contours in units of Vega magnitudes / square-arcsecond.  The approximate location and size of the occulting spot is marked with a white circle in each panel.  Note that the elongated spots located about 1$\arcsec$ from the center are purposefully induced  in the point spread function (called ``satellite spots'') in order to estimate the location and flux of the central source hidden behind the coronagraphic spot (see text for further explanation).  East is left, North is up.  }
\label{fig:toti}
\end{figure}

\begin{figure}
\centering
\includegraphics[width=4in]{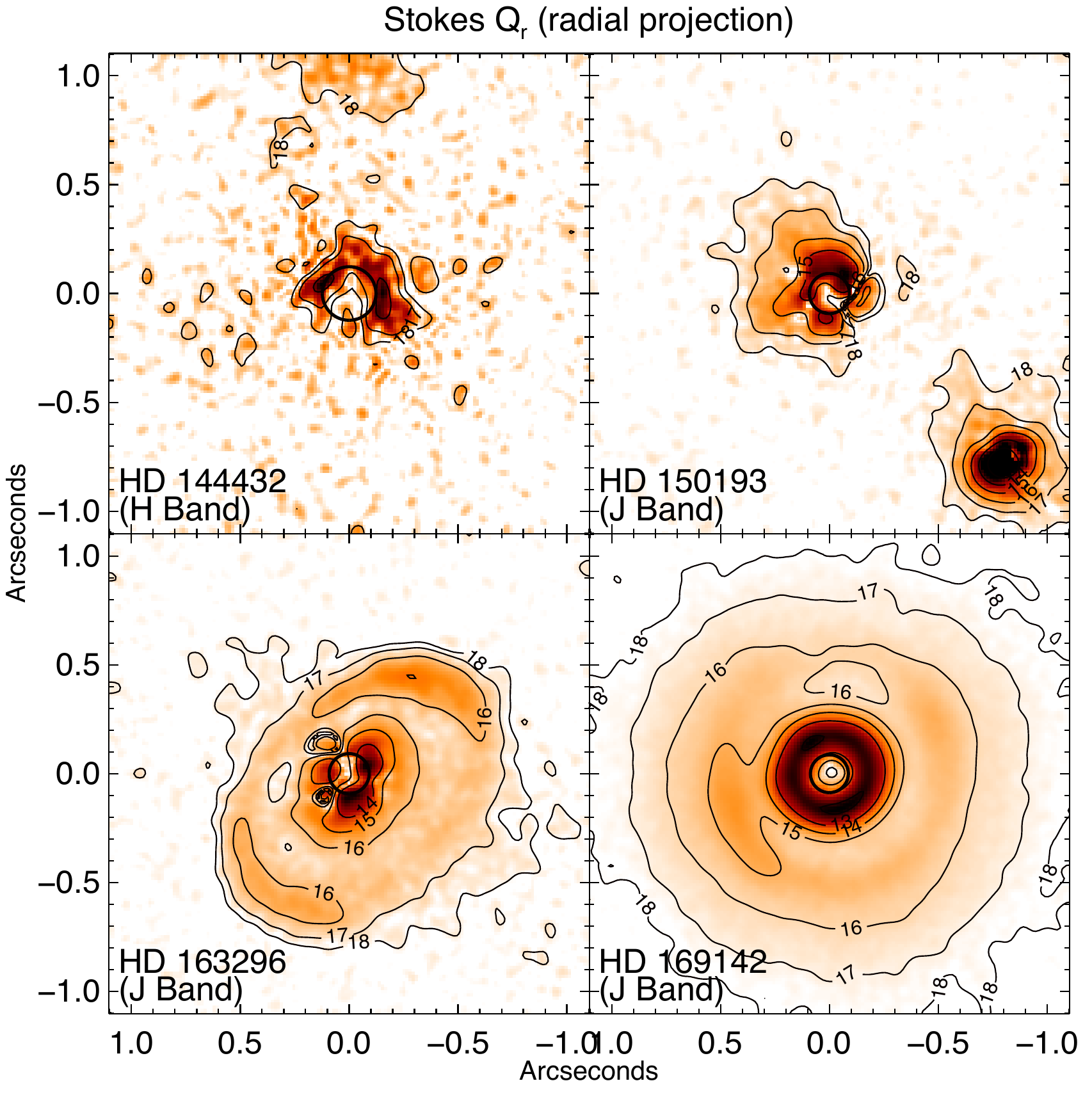}
\caption{Radially-projected $Q_r$ maps, measured using the Gemini Planet Imager, shown using a color table proportional to the square-root of absolute value of polarized intensity $Q_r$ (see text for description of this quantity).  The maps were smoothed by a Gaussian with FWHM 42~milliarcseconds (3~pixels) to improve SNR. The intensity scales are different for each panel and local surface brightness levels can be found as labeled contours in units of Vega magnitudes / square-arcsecond.  The approximate location and size of the occulting spot is marked with a black circle in each panel. East is left, North is up.  }
\label{fig:qr}
\end{figure}

\begin{figure}
\centering
\includegraphics[width=4in]{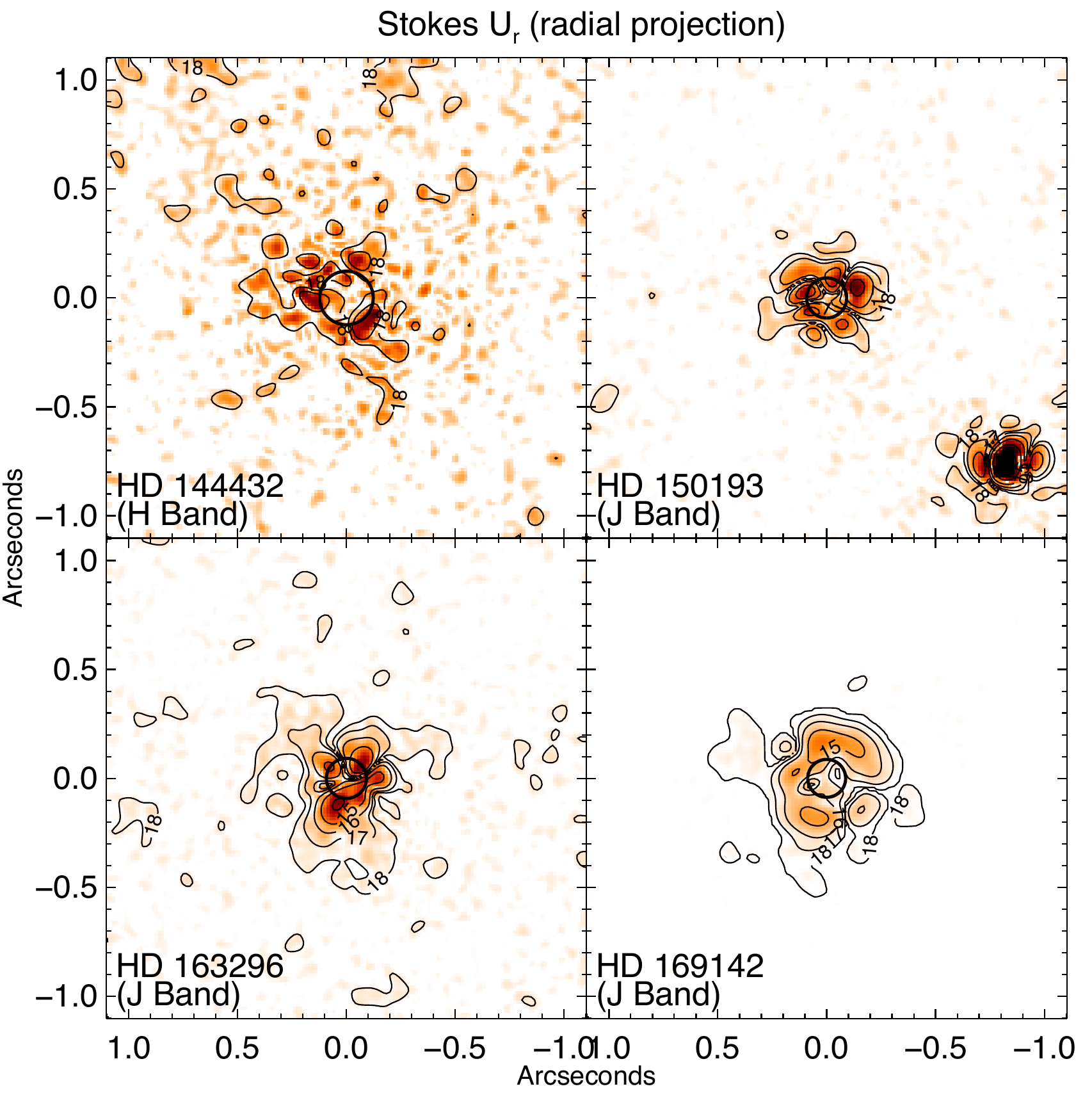}
\caption{Radially-projected $U_r$ maps, measured using the Gemini Planet Imager, shown using a color table proportional to the square-root of absolute value of polarized intensity $U_r$ (see text for description of this quantity).  The maps were smoothed by a Gaussian with FWHM 42~milliarcseconds (3~pixels). The intensity scales for each panel share the same color table as used for $Q_r$ in Figure~\ref{fig:qr} and  local surface brightness levels can be found as labeled contours in units of Vega magnitudes / square-arcsecond.  The approximate location and size of the occulting spot is marked with a black circle in each panel. East is left, North is up.  }
\label{fig:ur}
\end{figure}

\begin{figure}
\centering
\includegraphics[width=4in]{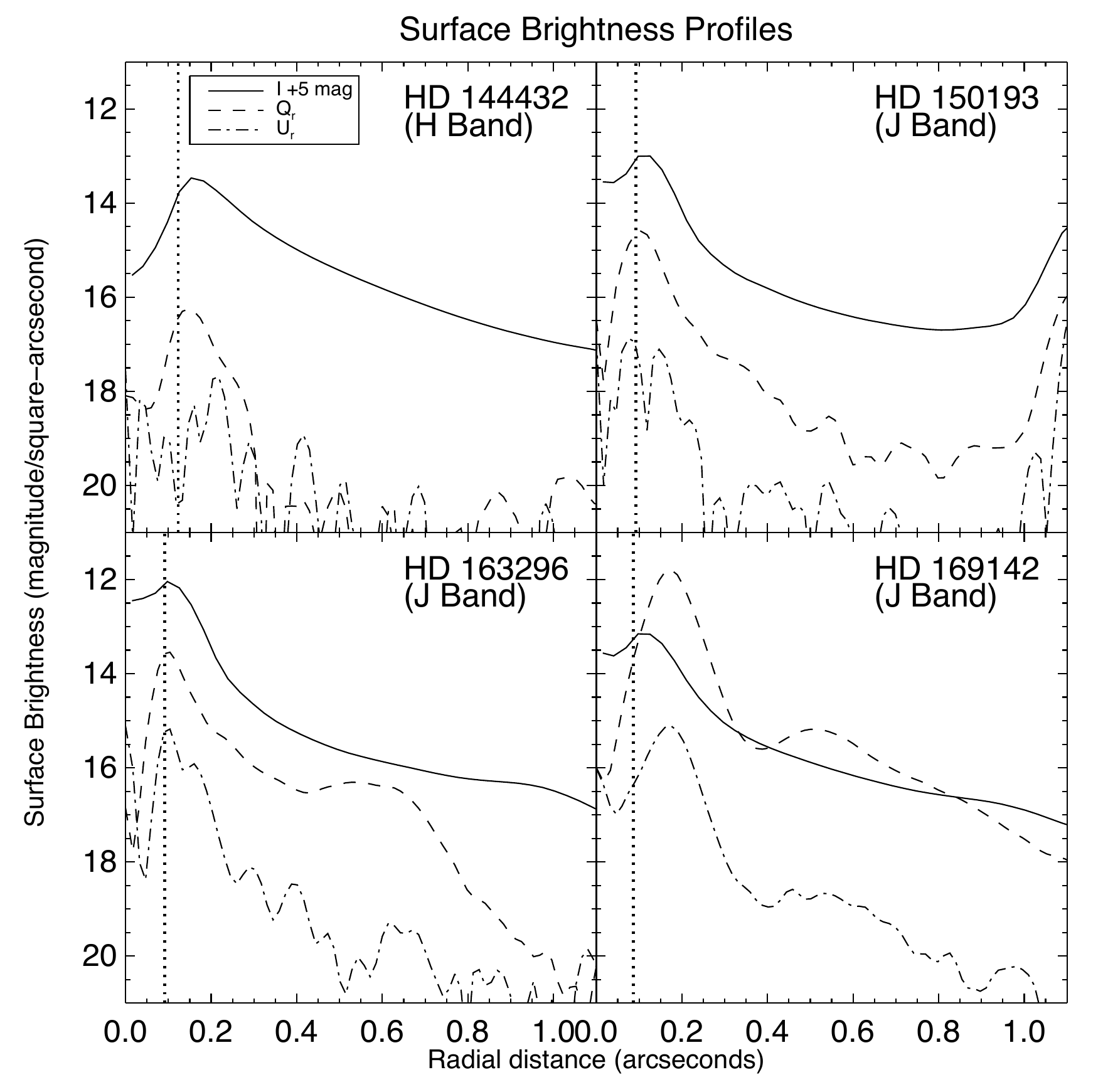}
\caption{Azimuthally-averaged surface brightness profiles for the four target stars.  After averaging in annuli centered on the stellar position, we smoothed by 42~milliarcseconds (3 pixels) and then took the absolute value before plotting the logarithm.  Note that the total intensity profile has been shifted by 5 magnitudes ($\times100$). \edit1{The dashed vertical line shows the radius of the coronagraphic spot used.  Note that the scatter in each curve give an estimate of the statistical errors remaining after the azimuthal averaging while the magnitude of the $U_r$ profile is representative of the radial-dependent systematic errors, since we normally expect a small or non-existent $U_r$ astrophysical signal.}}
\label{fig:profiles}
\end{figure}

\section{Basic Results}
We present the total intensity maps in Figure~\ref{fig:toti}, Stokes $Q_r$ maps in Figure~\ref{fig:qr}, Stokes $U_r$ maps in Figure~\ref{fig:ur}, and their corresponding mean radial profiles in Figure~\ref{fig:profiles}.  Each figure has an explanation of how the images are scaled and presented.  We generally present the absolute value of the $Q_r$ and $U_r$ maps with logarithmic surface brightness contours.  

First we discuss the total intensity maps in Figure~\ref{fig:toti}.  We see a depression in the center of the PSF because of the occulting mask, marked by a circle.  We see the PSF was rather elongated for the HD~144432 observation while more circularly-symmetric in the other observations \edit1{likely due to telescope wind-shake.}  One can see diffraction spikes from the $\sim$1.4\arcsec companion near the top of the HD~144432 frame and one can also see the slightly-saturated companion to HD~150193 located 1.12$\pm$0.02\arcsec at PA 223$\pm$1\arcdeg relative to the primary. The other spots are due to either residual ``waffle mode'' from the adaptive optics system or the diffractive satellite spots induced by GPI for registration of the bright star behind the coronagraph. We assume this light is from the central source PSF and that we can not extract the scattered light intensity from circumstellar dust without using the polarization signature.

The main results of this work are best seen in Figure~\ref{fig:qr}, where the radially-projected Stokes $Q_r$ maps are presented.  It is useful to compare these images to the $U_r$ images in Figure~\ref{fig:ur} since residuals in the $U_r$ map indicate the level of systematic errors in our suppression technique, recalling the caveat that we are not observing more edge-on systems \citep[as discussed by][]{canovas2015}.  We will discuss briefly each target separately.  

\begin{itemize}
\item HD~144432. \edit1{HD~144432 has been classified as a Herbig A9Ve star with a Meeus Group IIa, meaning that the mid-IR dust continuum is a power law with solid state features \citep{meeus2001}.  \citet{muller2011} reports that the 1.47'' companion is itself a close binary consisting of K7V and M1V stars. } We see no convincing sign of scattered light at H band since the level and pattern of $Q_r$ is similar to that seen in $U_r$.  While the surface-brightness limits within 0.25\arcsec~are only about $\sim$16.5 mag/square arcsecond, our upper limits beyond 0.3'' reach about 20 mag/square arcsecond.   We could not locate any previous measurements to compare.  
HD~144432 has amongst the smallest far-IR excess from our sample.  We suspect the presence of the nearby companion is responsible for truncating the outer disk, although this is difficult to prove without knowledge of the full orbit of HD~144432B.

\item HD~150193. \edit1{HD~150193 has been classified as a Herbig A1Ve star with a Meeus Group IIa  \citep{meeus2001}.  \citet{carmona2007} reports that the 1.1'' companion has a F9Ve spectral type. } There is slight excess $Q_r$ emission seen to the North-East beyond about 0.2$\arcsec$ with surface brightness of $\sim$17 mag/square arcsecond.  This pattern is not seen in the $U_r$ emission and appears to be real, although some caution is advised since this source has by the far highest intrinsic stellar polarization ($\sim$3\%) and thus calibration errors may be larger than for other targets.  \citet{garufi2014} reported an upper limit of $\sim$0.8 mJy/square arcsecond at this radius at H band which corresponds to about 15 mag/square arcsecond, not incompatible with our new detection even accounting for the $J-H=0.7$ for this star.  We note that HD~150193 also has small far-IR excess like HD~144432, quite possibly due to the disk-disrupting presence of stellar companions in both systems.  \edit1{Note that the very low scattering flux could also be due to the geometrical effect of self-shadowing (i.e., being a Meeus Group II object), a fact pointed out recently by \citep{garufi2014}.}

\item HD~163296.  \edit1{HD~163296 has been classified as a Herbig A3Ve star with a Meeus Group IIa  \citep{meeus2001}.  There are no known companions to this source.} In the $Q_r$ image, we clearly detect a full outer ring (not centered on the star) with major axis of 0.65\arcsec~oriented along PA of $\sim$136\arcdeg~with peak polarized intensity of $\sim$15 mag/square arcsecond, confirming the lower SNR detection by \citet{garufi2014} at H and K. Only with our new high-SNR GPI data can we clearly see the ring is off-center.  Within 0.4\arcsec, there does appear to be excess $Q_r$ emission in the inner region beyond the residual seen in $U_r$.  This excess is elongated the same way as the outer ring giving further evidence that we are indeed seeing scattered light from dust within the inner disk.   \citet{garufi2014} found the peak K band (H band) surface brightness of the outer ring to be 0.5 (0.15) mJy/square arcsecond which is $\sim$15.2 (17.1) mag/square arcsecond.  \edit1{As this disk is known to be highly inclined \citep[$\sim$48\arcdeg,][]{tannirkulam2008}, there may be scattered light showing up in the $U_r$ image due to multiple scattering \citep{canovas2015} but a detailed discussion of this will require a full radiative transfer calculation that is beyond the scope of this work.  However, even without a full model, we can analyze the geometry of this J band image using basic arguments  } and we offer a simple interpretation for the off-center ring in \S\ref{mwc275}.

\item HD~169142. \edit1{HD~169142 has been classified as a Herbig A5Ve star with a Meeus Group Ib, meaning that the mid-IR dust continuum is a power law plus a black body with no solid state features \citep{meeus2001}. This is only group I object in our sample.  There are no known companions to this source.} In the $Q_r$ image, we clearly detect the inner ring and outer disk first reported  by \citet{quanz2013} at H band. Our new J band image is much higher quality and more comparable to the recent Subaru polarization imaging by \citet{momose2015} at H band; the peak of the inner ring is well distinguished from the coronagraphic spot (see profiles in Figure~\ref{fig:profiles} in our data).   We also see a gap in the outer disk to the north (see Figure~\ref{fig:qr}). \cite{momose2015} found peak surface brightness level of the outer disk to be 4 mJy/square arcsecond, or 13.5 mag/square arcsecond.  At J band, we find a peak surface brightness about $\sim$14.5 mag/square arcsecond.  Note that the inner edge of the ring is not so clearly resolved by \citet{momose2015} as it is for our work and \citet{quanz2013}.  Specifically we find the inner ring has a sharp inner edge with radius of 0.18\arcsec (25~AU), the outer disk has peak surface brightness at 0.51\arcsec (75~AU), with a local dip in brightness at 0.38\arcsec (55~AU) -- \edit2{see Figure\,\ref{fig:azimuth} for detailed surface brightness curves}.  The \edit1{mean} J-H color of the inner ring in our image is $\sim$0.4, similar to the color of the star itself, while the outer disk has a mean color J-H$\sim$1.0, much redder.  We describe our attempts to model the disk of HD~169142 in the next section.

\end{itemize}

\section{Radiative transfer modeling of HD~169142}
\label{model}
We used the  {\sc torus} radiative transfer code \citep{harries_2000,harries_2004,harries_2011} to model HD~169142. The code uses the Monte Carlo (MC) radiative equilibrium method of \citet{lucy_1999}, and an adaptive mesh that is constructed to adequately resolve sharp opacity gradients. The {\sc torus} code has been extensively benchmarked \citep{harries_2004, pinte_2009}. In the following section we describe the  properties of the circumstellar disk model and then detail the optimization process used to identify the best fit. Finally we critically appraise our model solution.

\subsection{The disk structure}

The disk density in cylindrical coordinates $(r,z)$ is given by
\begin{equation}
\rho(r,z) = \rho_0 \left( \frac{r}{r_0} \right)^{-\alpha} \exp \left( -\frac{1}{2} \frac{z^2}{h(r)^2} \right)
\label{density_eq}
\end{equation}
where $\rho_0$ is a fiducial density, $\alpha$ is the density power-law index, and the scale-height $h(r)$ is given by 
\begin{equation}
h(r) = h_0 (r / r')^\beta
\label{scaleheight_eq}
\end{equation}
where $h_0$ is the scale-height at $r'$ and $\beta$ is the flaring index. The value of $\rho_0$ is found from
\begin{equation}
M_{\rm disk} = \int_{R_{\rm i}}^{R_{\rm o}} \int_{-\infty}^{\infty} 2\pi r\rho(r,z)\,dz\,dr 
\end{equation}
where $R_{\rm i}$ and $R_{\rm o}$ are the inner and outer disk radii respectively.

Previous authors \citep{quanz2013,osorio_2014,wagner2015,momose2015,seok2016} have identified the main components of the disk from analyses of the SED and scattered light imaging, namely: (i) a small ``inner disk'' that provides the near-IR excess , (ii) a gap that is apparently very low density between the innermost disk and $\sim 20$\,AU, a (iii) a narrow ring of material between $\sim 20$\,AU and $\sim 40$\,AU (we refer here to this as the ``inner ring''), (iv) a second gap, with a reduced (but non-negligible) density, spanning the region between $\sim 40$\,AU and $\sim 70$\,AU, and finally (v) a flared, ``outer disk'' extending from $\sim 70$\,AU out to around 245\,AU.   \edit1{For the rest of this paper we will consistently refer to the three main regions as the ``inner disk'' which is unresolved by our AO observations, the``inner ring'' which we clearly resolve here with GPI, and the ``outer disk'' which we also see in this paper in polarized light.}

\begin{figure}
\centering
\includegraphics[width=4in]{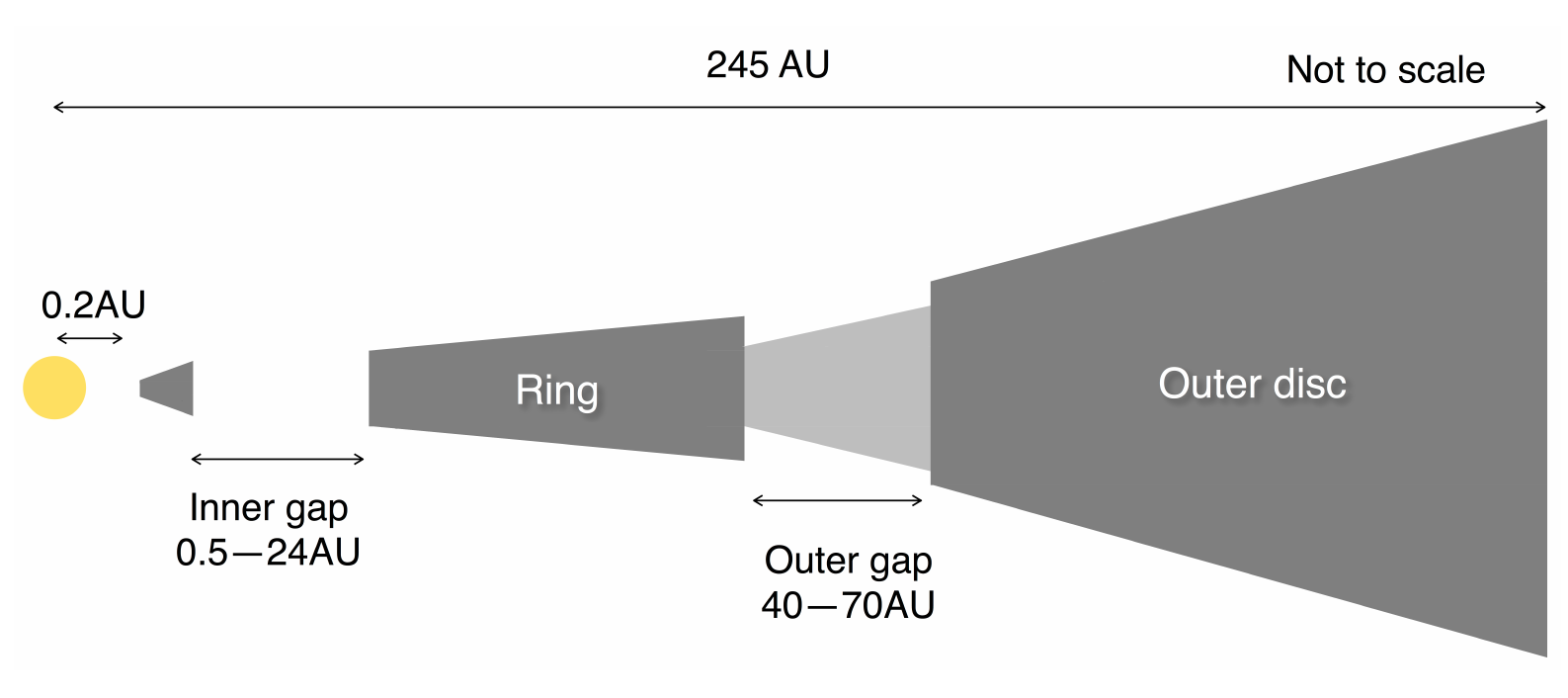}
\caption{This figure illustrates the basic components of our disk model for HD~169142, with detailed parameters listed in Table~\ref{params_tab} . }
\label{fig:cartoon}
\end{figure}

Much of the radiation observed results from heating of the inner edge of the disk, and outer walls of the gap by stellar radiation, whilst the mid-planes of the walls at $\sim 20$ and $\sim 70$\,AU are shadowed by the inner disk and the ring respectively. The complex nature of this interaction leads to substantial degeneracies between the free parameters of the model, for example increasing the scale-height at $\sim 20$\,AU or reducing the wall scale-height at 0.2\,AU will lead to an increase in the mid-IR flux, since both changes lead to a greater area exposed to direct stellar heating (although the latter change will lead to a reduction in the near-IR excess).  Naturally this complicated disk model means that there are many free parameters (see Table~\ref{params_tab}), and our initial aim was to fix as many of these at plausible values prior to a formal best-fit procedure.

\subsection{Determining the best fit model}

We adopted the distance to HD~169142 of 145\,pc from \citet{sylvester1996} and the stellar parameters from \citet{manoj2006}. We used a Kurucz model atmosphere of appropriate $T_{\rm eff}$ and $\log g$ to represent the photospheric emission. See Table~\ref{params_tab} for full set of assumptions.

For the dust,  \cite{draine_1984} silicate grains were used for our disk model, with grain size populations described by the usual power-law distribution \citep{mrn1977}
\begin{equation}
n(a) \propto a^{-3.5}
\end{equation}
where $n(a)$  is the number density of grains with a size $a$. We used two distinct grain-size populations (small and large) whose properties are listed in Table~\ref{params_tab}. Since there is no obvious 10\,$\mu$m silicate feature in the SED, we concluded that the \edit1{inner disk} consists of large grains only, whereas over the rest of the disk we invoke grain settling, with the large grains having a scale-height that is half the local gas scale-height, and the \edit1{grain mass fractions are chosen to give a constant dust-to-gas ratio of 1\% throughout the disk}. Thus the midplane of the disk is dominated by large grains, and the scattering region of the upper disk consists of small grains.

The location of the inner disk was fixed at 0.2\,AU, and the disk scale-height at that radius was a free-parameter of the model (the varying of which allows a fit to the near-IR excess in the SED). The locations of the two gap walls is quite well determined from the peaks in the polarized surface brightness profiles at $J$ and $H$, and our preliminary modeling, and we fixed these at 24\,AU and 70\,AU. The radius of the inner gap is not strongly constrained and we fixed this at 0.5\,AU, whilst the inner radius of the outer gap is quite well constrained by the polarized surface brightness profiles and we fixed this at 40\,AU. The outer disk radius was fixed at 245\,AU. The disk mass and flaring index $\beta$ are quite strongly constrained by the far-IR and millimeter data points in the SED, and we found that a disk gas mass of 0.05$M_\odot$ and a flaring parameter of 1.09 gave a good fit. We choose to use a canonical surface density fall off of $\Sigma(r) \propto r^{-1}$ following \citet{momose2015}.

The inner gap density was set to the floor density of the radiation transfer calculation ($10^{-24}$\,g\,cm$^{-3}$), however the outer gap is clearly visible in both scattered light images, and therefore must have a non-negligible density. We therefore chose to parameterize the density in the gap ($\rho_{\rm gap}$) by a simple scaling factor $f$ where
\begin{equation}
\rho_{\rm gap}(r,z) = f \rho(r,z)
\end{equation}
where $\rho(r,z)$ is the density calculated from Equation~\ref{density_eq} and Equation~\ref{scaleheight_eq} using the scale-height at 70\,AU, and $0 < f \leq 1$.

The free parameters in the model are now the scale-heights of the inner disk, the ring,  and the outer disk, the density in the outer gap, and the flaring parameter $\beta$. We wished to determine the best fit in the most objective way possible, and therefore we chose to fit the $J$ and $H$ polarized intensity profiles\footnote{H band images from \citet{quanz2013} were kindly provided by Dr. Quanz.} and the SED simultaneously.  A cartoon schematic of our disk mode can be found in Figure~\ref{fig:cartoon}.

After fixing some parameters as discussed above, we used a generic algorithm to search the remaining parameter space. We modified the {\sc pikaia} routine by \cite{charbonneau_1995} to call a bespoke (i.e., custom) goodness-of-fit function. This function creates a {\sc torus} input deck containing the fixed and free parameters, and runs {\sc torus} in a parallel mode for speed. Once {\sc torus} has completed, the function generates $J$ and $H$ band azimuthally-averaged polarized intensity profiles from the appropriate images, and also reads in the simulated SED. Finally the routine calculates a reduced $\chi^2$ value by comparison with the data and returns the inverse of this ({\sc pikaia} expects a goodness-of-fit value that is numerically larger the better the fit). We operated {\sc pikaia} in a full-generational replacement mode with 100 individuals per generation and searched parameter space for 50 generations. \edit1{We do not quote formal errors on our best fit quantites, due to a combination of systematic uncertainties on our polarimetry and degeneracies between our model parameters. Furthermore although the generic algorithm has a mutation probability in order to avoid convergence to a local minimum in the chi-squared hypersurface, it is possible other solutions of comparable goodness-of-fit exist.}

\begin{table}
\caption{Model parameters for HD169142}
\label{params_tab}
\begin{tabular}{lll}
\hline
Parameter & Value & Description \\
\hline
\multicolumn{3}{c}{Stellar parameters} \\
Stellar radius, $R_*$ &  2.2\,R$_\odot$  & \citet{manoj2006}\\
Effective temperature, $T_{\rm eff}$ & 8100\,K & \citet{manoj2006} \\
Stellar mass, $M_*$ & 2\,M$_\odot$ & \citet{manoj2006}\\
Distance                  & 145\,pc & \citet{sylvester1996} \\
$A_v$ & 0.5 & \citet{dent2006} \\
\multicolumn{3}{c}{Disk parameters} \\
Inclination, $i$ & 13$^\circ$ & \citet{panic2008} \\
Disk mass, $M_{\rm disk}$ & 0.05\,M$_\odot$ & Refined\\
Disk flaring index, $\beta$ & 1.09 & Refined\\
Radial density index, $\alpha$ & $-$2.09 & Refined\\
Inner disk radius, $R_{\rm i}$ & 0.2 \,AU & Fixed \\
Scale-height at inner radius & 0.006\,AU & Fitted \\
Outer disk radius, $R_{\rm o}$ & 245\,AU & Fixed \\
Inner disk gap range  & 0.5--24\,AU & Refined \\
Scale-height at 24\,AU & 1.14\,AU & Fitted \\
Inner gap density & $10^{-24}$\,g\,cm$^{-3}$ & Fixed \\ 
Outer disk gap range & 40--70\,AU & Fixed \\
Scale-height at 40\,AU & 4.06\,AU & Fitted \\
Outer gap density scaling ($f$) & 0.15 & Fitted \\

\multicolumn{3}{c}{Grain properties, small grains} \\
Grain type    & Silicates & \citet{draine_1984} \\
Min grain size, $a_{\rm min}$ & 0.01\,$\mu$m & Fixed \\
Max grain size, $a_{\rm max}$ & 1\,$\mu$m & Fixed \\

\multicolumn{3}{c}{Grain properties, large grains} \\
Grain type    & Silicates & \citet{draine_1984} \\
Min grain size, $a_{\rm min}$ & 5\,$\mu$m & Fixed \\
Max grain size, $a_{\rm max}$ & 1000 \,$\mu$m & Fixed \\

\hline
\end{tabular}
\end{table}

\subsection{Results}

\begin{figure}
\centering
\includegraphics[width=4in]{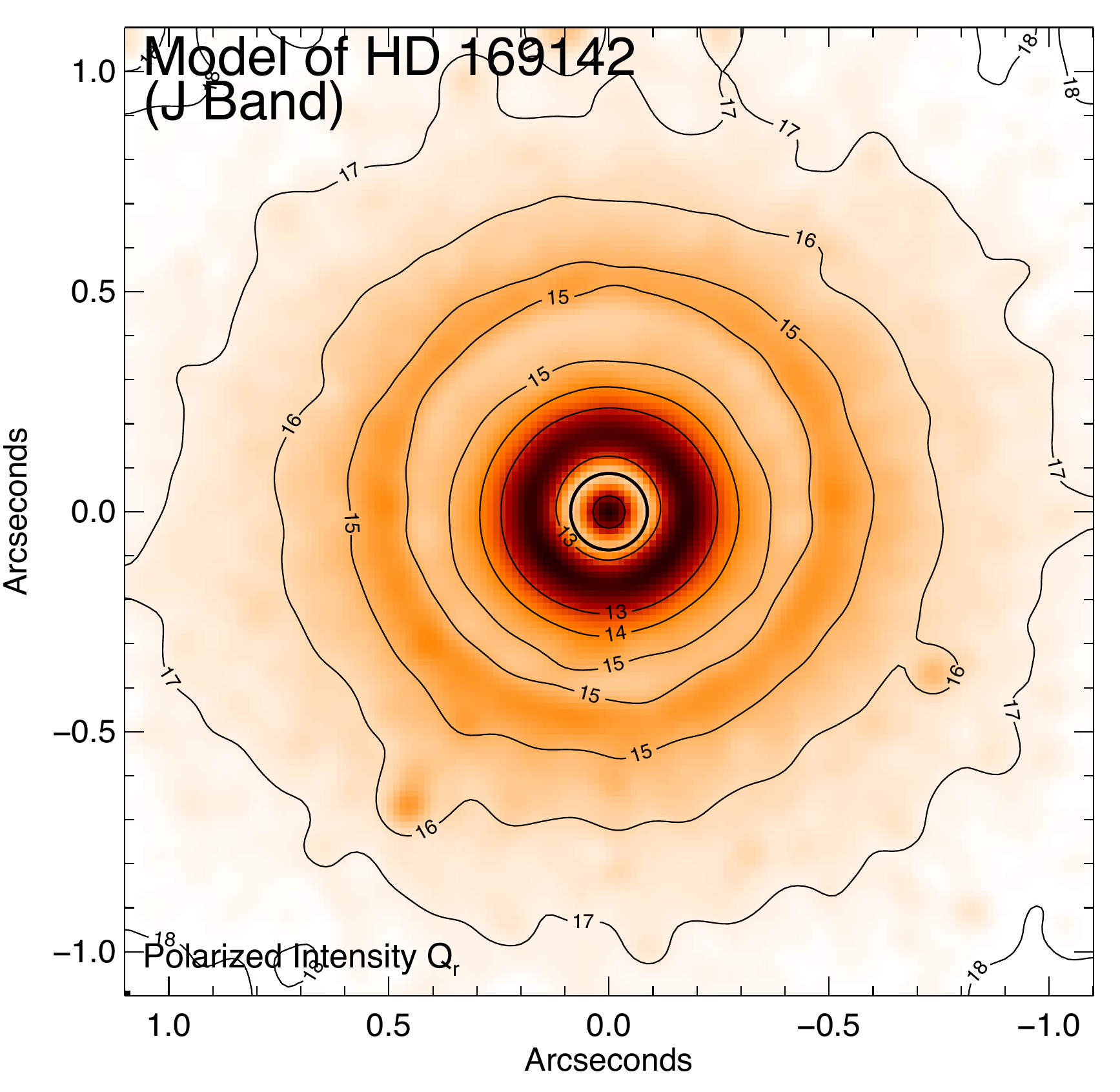}
\caption{Simulated image of HD~69142 based on the radiative transfer model of the polarized intensity at J band, smoothed to same angular resolution as our data with identical color table found in Figure~\ref{fig:qr}.  Note that some of the irregularity in the outer disk arises from stochastic ``noise'' in the Monte Carlo sampling during the radiative transfer calculation.}
\label{fig:model_qr}
\end{figure}

\begin{figure}
\centering
\includegraphics[height=2.5in]{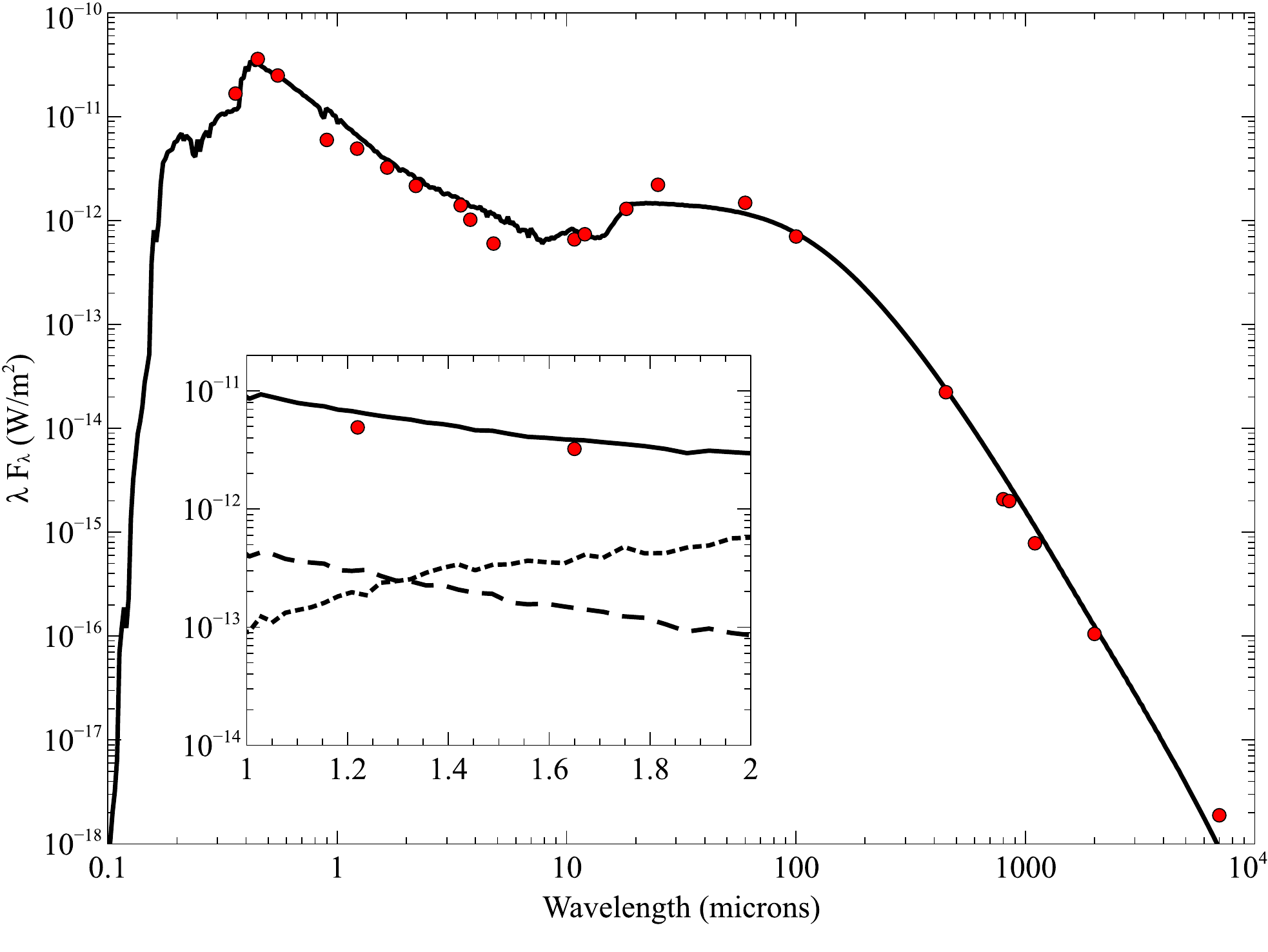}
\includegraphics[height=2.5in]{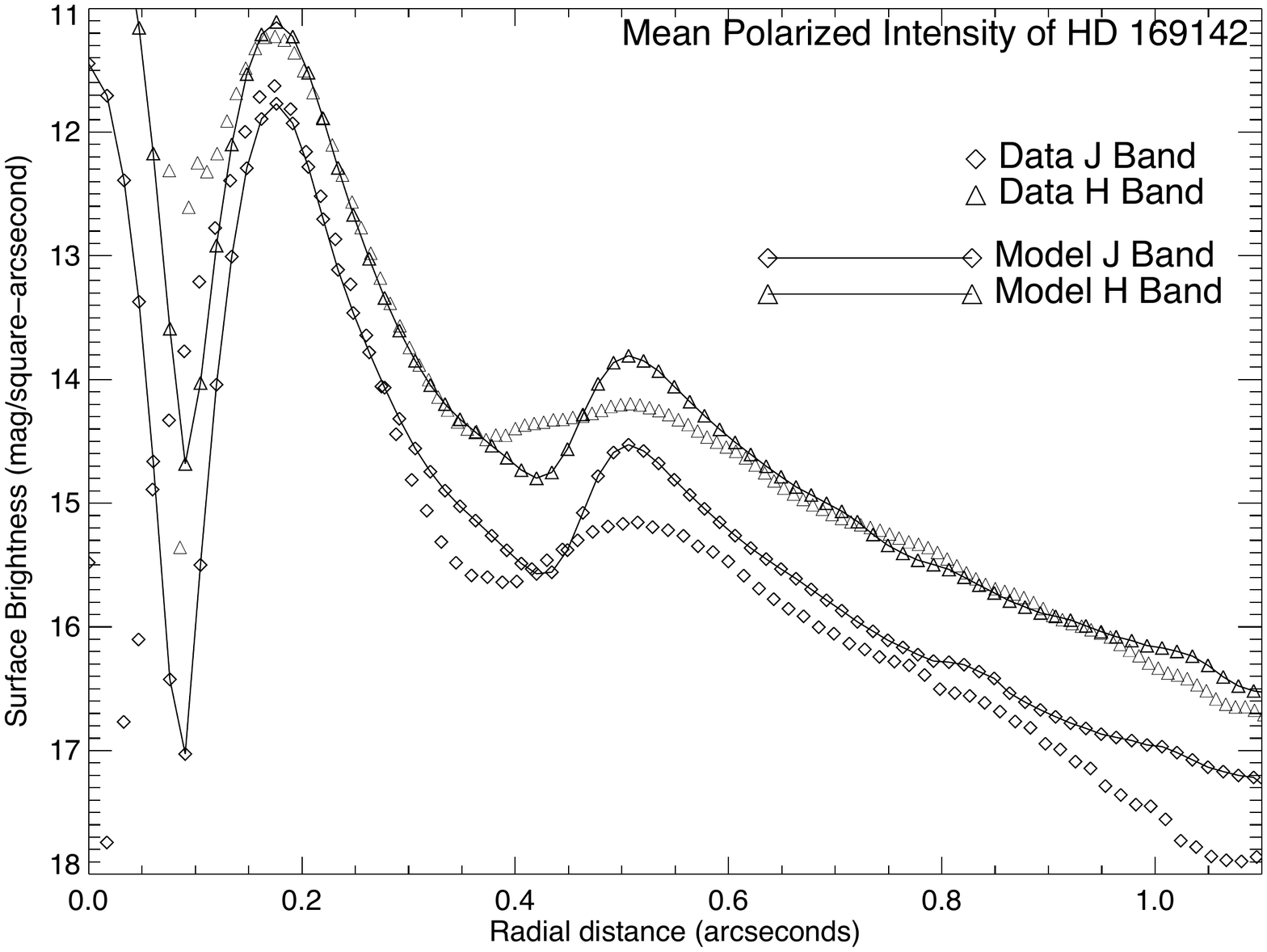}
\caption{
\edit2{Model-fitting results for HD~169142 (see \S\ref{model} for full details). (left panel) Comparison of the input SED and model fit for HD~169142 radiative transfer model discussed in \S\ref{model} using photometry collated by \citet{dent2006}.  The inset figure shows a zoom-in of the region around the J and H bands, with the two photometric points as red circles and the model SED as a solid line. The contribution to the SED from scattered protostellar light (dashed line) and the scattered thermal emission (dotted line) are shown. }
(right panel) Comparison of mean surface brightness profiles of the polarized intensity and J and H band.}
\label{fig:model_profile}
\end{figure}

The best fit parameters are listed in Table~\ref{params_tab} and a model image of the polarized intensity at J band can be found in Figure~\ref{fig:model_qr}. One can see the major features of the GPI image are reproduced by comparing to the data presented in Figure~\ref{fig:qr}.
The agreement with the SED is also satisfactory (see left panel Figure~\ref{fig:model_profile}). The fit is poorest at around 60\,$\mu$m. We note that this part of the SED arises primarily from the inner wall of the outermost part of the disk (corresponding to the second maximum in the surface brightness profiles at 0.55\arcsec). A better fit to the SED at this point is possible by changing the scale height of the inner ring (in order that the wall sees more direct stellar flux) but this would compromise the surface brightness profile fit. \edit1{We determine a disc mass (gas and dust) of 0.05\,M$_\odot$ (or equivalently a dust mass of $5 \times 10^{-4}$\,M$_\odot$) which is smaller than that of 0.12\,M$_{\odot}$ found by \citet{osorio_2014} but larger than that of \citet{maaskant2013}, who obtained $M_{\rm dust}=0.8 \times 10^{-4}$\,M$_\odot$.}

We also show a detailed comparison of fit to the polarized surface brightness profiles at J and H band in the right panel of Figure~\ref{fig:model_profile}.  These profiles are an adequate match to the observations  in terms of both the inner wall of the ring (the peak in the surface brightness at around 0.2\arcsec) and in the gap (the minimum at around 0.35\arcsec) although improvement could be made with further model complications. Note that the gap is not completely empty ($f=0.15$) as first noted by \citet{quanz2013} and confirmed through the modelling of \citet{wagner2015}. For scenarios in which the gap is 
produced by a companion, the degree of clearing, or equivalently, the density  contrast between the gap and the unperturbed disk, will be a function of both 
disk properties and companion properties. This dependence is complicated, 
but can provide constraints on possible companions, and should be pursued
further in future work.

We find the power-law of the brightness profiles out to 1\arcsec~ is well matched by the model, and strongly constrains the flaring power law index ($\beta\sim1.09$). The relative brightness of the $J$ and $H$ profiles in the model shows reasonable agreement with the data, although the amount of differential reddening between the two rings is not fully reproduced.

If one only considers the polarized intensity, the observed reddening of the outer disk could be interpreted in terms of a size-dependent scattering coefficient.  We expect small grains ($a_{\rm max}\sim$0.1$\mu$m) to behave like Rayleigh scatterers and be relatively blue when looking at the near-infrared J,H,K bands.  However, for distributions with some medium-sized grains ($a_{\rm max}\sim$1$\mu$m), we start to see lower polarization fraction for $J$ band compared to $H$ and $K$ bands.  Thus one could qualitatively interpret our observed reddening as the presence of larger grains in the outer disk and small grains in the inner ring.  

\edit2{
In exploring the possibilities of different grain populations throughout the disk, we did a careful comparison of our model with the disk model by \citet[][hereafter SL16]{seok2016}.  We note that the 10$\mu$m emission seems to have comparable contributions from the ``inner disk'' component and the ``inner ring'' component  (see Figure 1 in SL16) which means that one can have weak silicate feature emission in one or the other region and not notice it in the SED.  Indeed, SL16 has weak silicate emission from small grains in the inner disk while our model has weak silicate feature emission from the inner ring.  Given that SL16 did not fit the polarized intensities as we have, it is not clear the best solution and this points to the need for a comprehensive modelling effort in a future paper.  For now, we must be circumspect and not overgeneralize our interpretations based on our preliminary model.
}

\edit2{
While acknowledging the ambiguities, we point out one solution to reddening the outer disk in polarized light which is to simply redden the inner disk light  as it traverses the upper layers of the inner ring before scattering off the outer disk.  While our model does not reproduce the large magnitude of this effect with the standard disk prescriptions (see Figure~\ref{fig:model_profile} for quantitative comparison), future modeling efforts can explore a more sophisticated set of models as well as more complex types of dust that could have polarization properties with a more complex wavelength-dependence.  It came to our attention late in the modelling process that icy mantles on cold grains can also affect the polarization properties, and specific ices such as irradiated methanol ice could have a red reflection spectrum\textit{} \citep[see review by][]{brown2012}.
}

\begin{figure}
\centering
\includegraphics[height=2.5in]{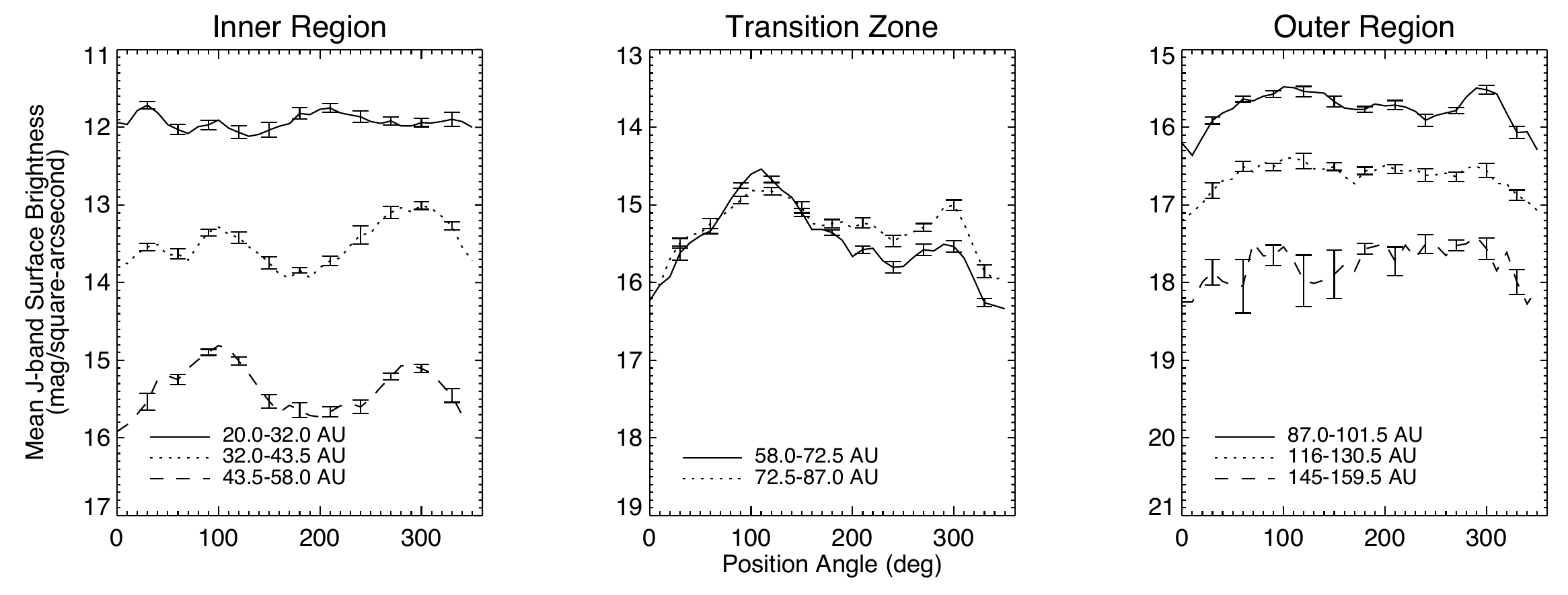}
\caption{Here we plot mean J-band surface brightness profiles of Polarized Intensity ($Q_r$) as a function of azimuth for certain deprojected annulli, based on the regions identified in \citet{momose2015}. We have assumed a disk deprojection following the prescription of \citet{momose2015}, specifically using a disk major axis PA of 5\arcdeg and disk inclination angle of 13\arcdeg. These profiles can be compared directly against the H-band profiles in that paper (as shown in their Figure 3).}
\label{fig:azimuth}
\end{figure}

\edit1{In order to facilitate quantitative comparisons between our J-band polarized intensity images and the the recent H-band images from \citet{momose2015}, Figure~\ref{fig:azimuth} shows mean surface brightness curves in various annuli as a function of azimuth.  We have used the disk deprojection  prescription of \citet{momose2015}, specifically using a disk major axis PA of 5\arcdeg and disk inclination angle of 13\arcdeg. We have chosen annuli to best match the same ones used by these workers, although we have included the inner ring as a separate curve including emission between 20-32\,AU.  Note that errors have been included in this figure and are based on variations between separate observations and does not include an overall flux calibration error discussed earlier.  }

\edit1{Based on a comparison of the azimuthal variations, } we do not find evidence that the inhomogeneities  in the inner ring are causing radial shadowing that can explain the azimuthal variations in brightness around the outer disk.  In this paper, we will not speculate on whether the dip in brightness at 55AU can be related to the presence of a forming planet in this system but have focused on describing our new high-quality imaging data which shall be used in future detailed modeling efforts.


\section{Discussion of HD 163296 ring}
\label{mwc275}

\begin{figure}
\centering
\includegraphics[width=4in]{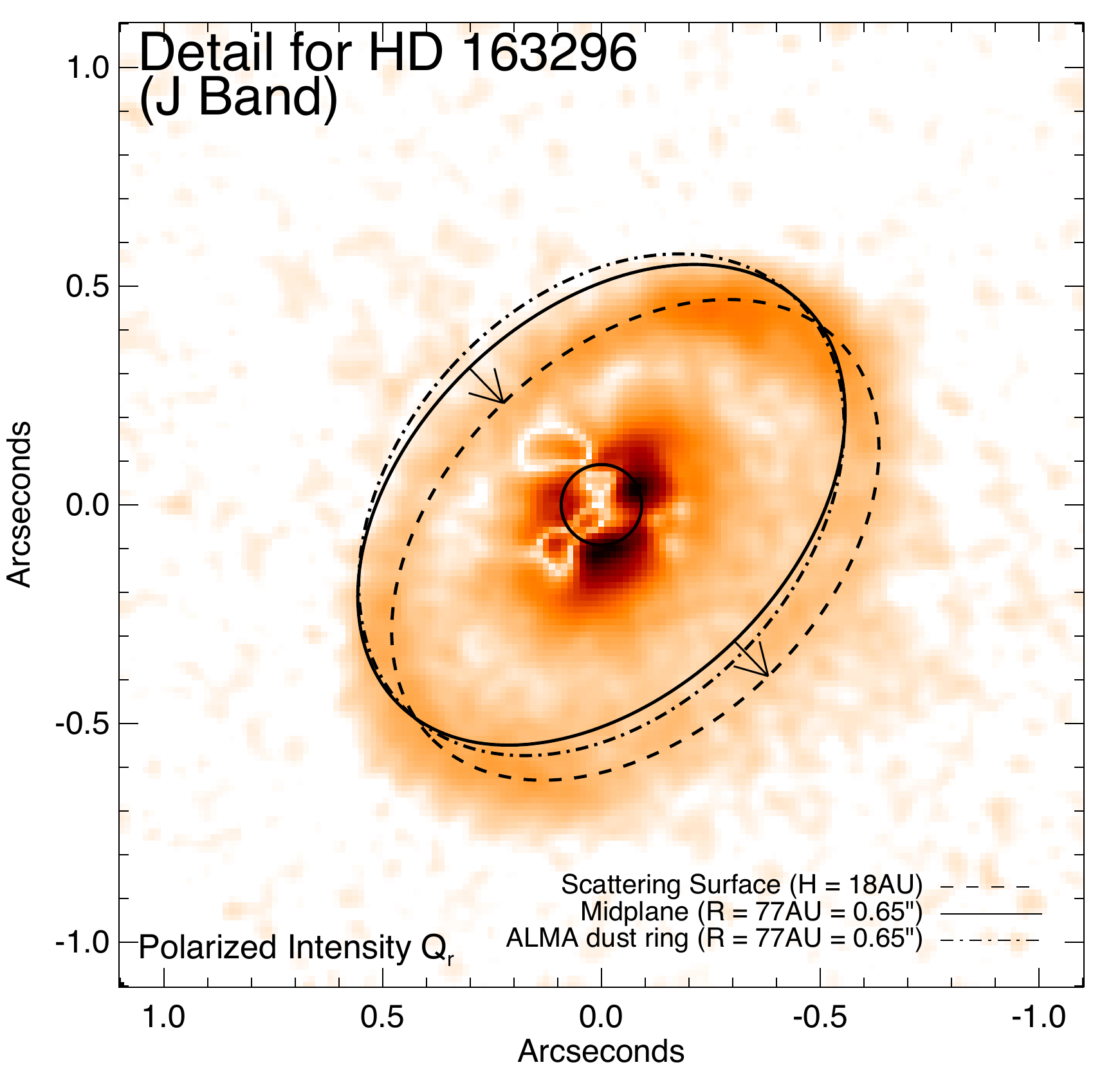}
\caption{Detailed analysis of HD~163296.  The solid line shows an ellipse centered on the star corresponding to inclination 48\arcdeg, PA 136\arcdeg \citep{tannirkulam2008} with radius 0.65\arcsec=77AU.  The off-center ring of emission can be explained as  by dust scattering  at a layer located 18 AU above the dust midplane (NE disk tilted toward the observer) and this layer is marked by a dashed line (arrows are included that point from the midplane up to the scattering layer, \edit1{marking the 0.1'' offset of the ring as projected on the sky}).  The dot-dashed line shows one of the 1.3mm continuum rings observed by ALMA \citep{zhang2016} -- indeed, the scattering we see at J band appears to be at the same location as the enhancement seen by ALMA possibly linking this dust enhancement to the location of the CO snow line at $\sim$90AU.  }
\label{fig:mwc275}
\end{figure}

\cite{garufi2014} first discovered evidence for an outer ring around HD~163296 but the data quality was marginal, \edit1{noting a possible ring offset ($\sim$0.05'') and } just clearly seeing the SE and NW lobes.  Here we see for the first time the full ring and can see the outline of the ring is strongly off-center (0.1'') from the star (see Figure~\ref{fig:mwc275}).  

While planets on eccentric orbits can produce off center debris disk rings \edit1{\citep[e.g., as discussed for Fomalhaut by][]{quillen2006},} we instead pursue the interpretation that dust scattering off of upper layers of an inclined, \edit1{flared} disk will cause the ring to appear off-center due to the viewing angle \edit1{\citep[see similar analysis by][]{deboer2016}}.  We first adopted the disk inclination and position angle measured by \citet{tannirkulam2008} using the CHARA interferometer observations of the inner disk (inclination 48\arcdeg, PA 136\arcdeg) and visually fit the radius of the ellipse $R$ and the height $H$ above the midplane for the scattering layer of dust.  We mark these ellipses for $R=0.65$\arcsec=77AU and $H=18$AU on the plot and find good agreement with the shape and orientation of the observed ring \edit1{(we can assign an approximate error of 2~AU to these values based on the angular resolution of our image). We compare the location of the scattering layer ( $\sim$18~AU) to the gas scale height which has been recently estimated to be 9~AU \citep{degregorio2013} and 6~AU \citep{guidi2016} based on modeling of mm-wave data. Indeed, we expect the dust scattering layer to be 3-6$\times$ higher than the gas scale height depending on wavelength and dust settling \citep[e.g.][]{dullemond2010}, and our results are qualitatively consistent with this. A future work should attempt to extract a wealth of quantitative information on the HD~163296 disk by modeling this off-center ring in more detail \citep[e.g.][]{stolker2016b} along with mm-wave ALMA imaging, but a full radiative transfer model is beyond the scope of this paper. We wish here to only approximately locate the radius and height of scattering material so we can compare to structures observed at other wavelengths.}

Recently, \citet{zhang2016} modeled ALMA 1.3mm data and suggested the presence of a local bright ring (one of three) of dust emission located at 0.65"(=77AU), just inside the CO snow line observed at $\sim$90AU.  \edit2{These results were confirmed by higher-resolution ALMA data just reported by \citet{isella2016} and we can see } that the small grains responsible for scattering near-infrared light at 18AU above the midplane are actually located directly above the larger grains observed in the mid-plane by ALMA.  If dust emission is enhanced just inside the CO snow line we see that both small and large grains are contributing.

This observation could have important consequences for modeling outer disk structures, especially when one considers that HD~169142 also shows an outer ring at a similar distance from the star -- although no measurement of its CO snow line has been made yet. We await high resolution ALMA images of both CO lines and mm-wave continuum to make these possible links between dust populations and snow lines more durable.

\section{Conclusions}

Here we have presented new, high signal-to-noise observations of HD~144432, HD~150193, HD~163296, and HD~169142 using polarized imaging of the Gemini Planet Imager.  We detected little or no scattered light around HD~144432 and HD~150193, two stars with nearby stellar companions that likely caused truncation of the outer disk.  GPI revealed an off-center ring around HD~163296 which can be understood as scattering off the upper layers of an outer dust ring that is also seen by ALMA.  Lastly, we contribute a new high-resolution image of HD~169142 that probes the structures of the two outer rings with unprecedented angular resolution and dynamic range, resulting in the first detection of strong differential color between the two rings.  The multiple rings seen for these last two objects, as well as other Herbig Ae/Be stars \edit1{such as AB Aur\citep{oppenheimer2008} and HD~97048\citep{ginski2016}},
need to be placed in the context of the expected ice lines. New ALMA CO imaging, linked with polarized scattered light imaging, should soon hopefully answer the question ``Are dust rings in the outer disks of Herbigs due to ice line chemistry, forming giant exoplanets, or something else?"

\acknowledgments
The authors would like to acknowledge productive conversations on GPI observing, data reduction and calibration techniques with 
 Fredrik Rantakyro, Pascale Hibon, Sascha Quanz, Douglas Brenner, and Max Millar-Blanchaer.   Also, the referee provided suggestions which made this work more comprehensive.
JDM/AA acknowledge support from NSF AST 12100972, 1311698, and 1445935.  We thanks Adam Rubenstein and Mike Sitko for help with photometry.   The calculations for this paper were performed on the University of Exeter Supercomputer, a DiRAC Facility jointly funded by STFC, the Large Facilities Capital fund of BIS, and the University of Exeter, and on the Complexity DiRAC Facility jointly funded by STFC and the Large Facilities Capital Fund of BIS. TJH acknowledges funding from Exeter's STFC Consolidated Grant (ST/J001627/1). SK acknowledges support from an STFC Rutherford Fellowship (ST/J004030/1) and a European Research Council (ERC) Starting Grant (Grant agreement No 639889).

This publication makes use of data products from the Two Micron All Sky Survey, which is a joint project of the University of Massachusetts and the Infrared Processing and Analysis Center/California Institute of Technology, funded by the National Aeronautics and Space Administration and the National Science Foundation. Based on observations obtained at the Gemini Observatory (programs GS-2014A-SV-412, GS-2015A-Q-49), which is operated by the Association of Universities for Research in Astronomy, Inc., under a cooperative agreement with the NSF on behalf of the Gemini partnership: the National Science Foundation (United States), the National Research Council (Canada), CONICYT (Chile), Ministerio de Ciencia, Tecnología e Innovación Productiva (Argentina), and Ministério da Ciência, Tecnologia e Inovação (Brazil).



\vspace{5mm}
\facilities{Gemini:South (GPI)}

\software{IDL,TORUS, GPI Data Reduction Pipeline\footnote{http://ascl.net/1411.018}}



\appendix
\section{Additional Figures}
\begin{figure}
\centering
\includegraphics[height=3.in]{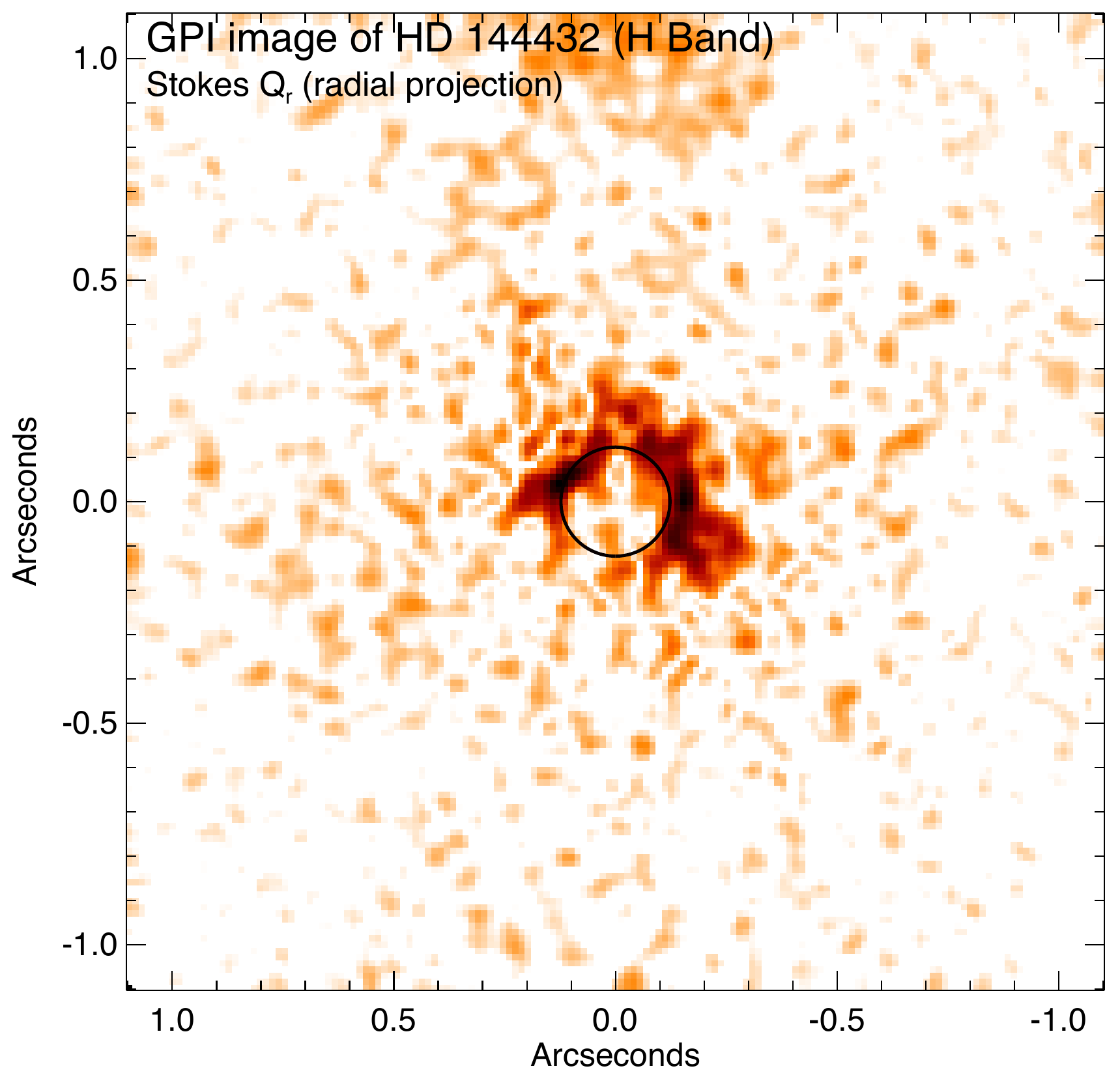}
\hphantom{........}
\includegraphics[height=3.in]{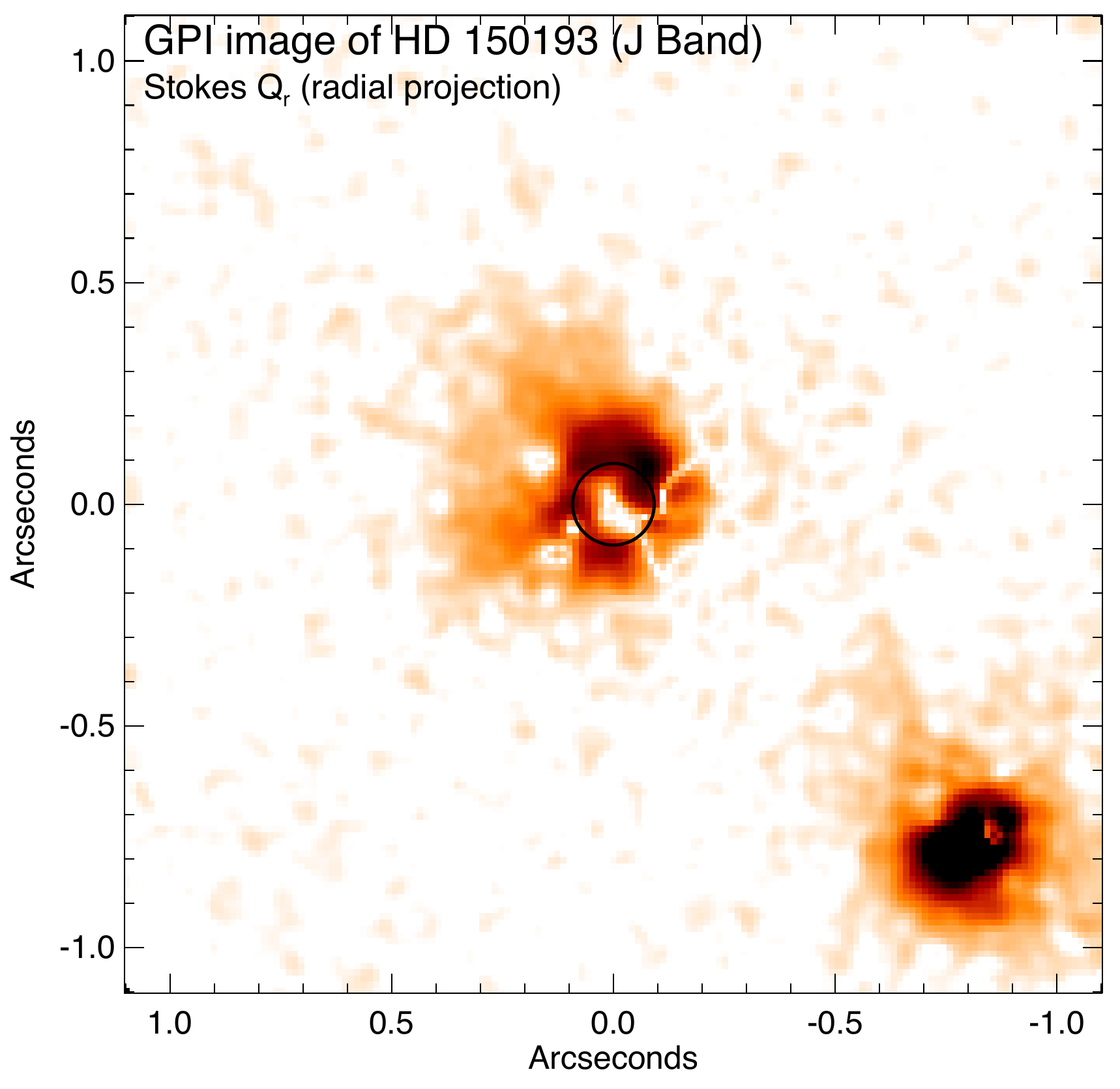}
\caption{The polarized intensity $Q_r$ images of (left panel) HD~144432 in H band and (right panel) HD~150193 in J band. We present these images without contours  to aid researchers in comparing our results with multi-wavelength imaging data from other facilities -- see Figure~\ref{fig:qr} for full details on the color table. \edit1{The circle marks the location and size of the coronagraphic spot used.}} \label{niceimages1}
\end{figure}

\begin{figure}
\centering
\includegraphics[height=3.in]{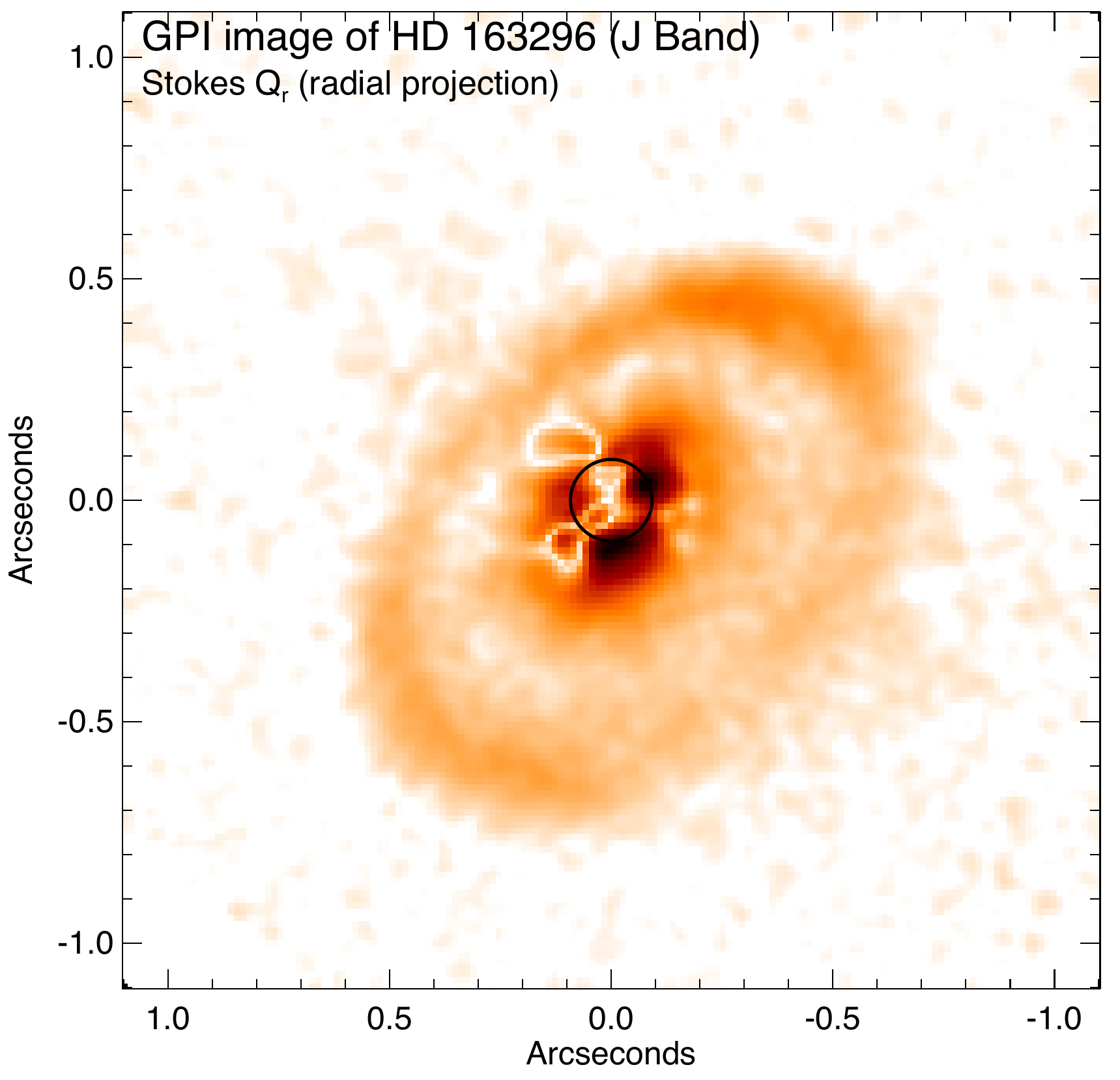}
\hphantom{........}
\includegraphics[height=3.in]{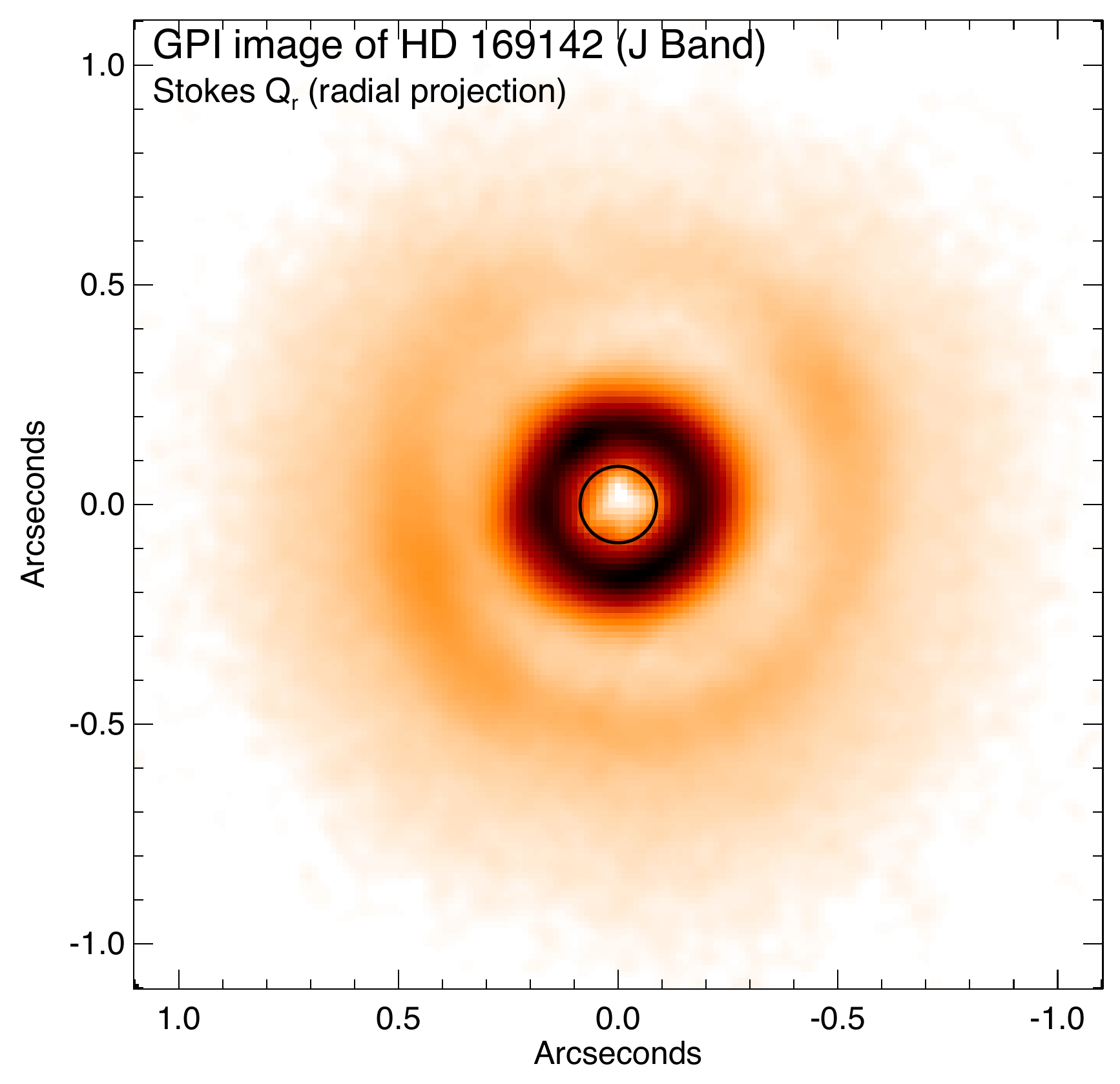}
\caption{The polarized intensity $Q_r$ images of (left panel) HD~163296 in J band and (right panel) HD~169142 in J band. We present these images without contours  to aid researchers in comparing our results with multi-wavelength imaging data from other facilities -- see Figure~\ref{fig:qr} for full details on the color table. \edit1{The circle marks the location and size of the coronagraphic spot used.} } \label{niceimages2}
\end{figure}

In order to aid other researchers in comparing our results to images taken at other wavelengths, we provide  reference figures (see Figures~\ref{niceimages1} \& \ref{niceimages2}) here of our polarized intensity $Q_r$ surface brightness maps without contours or distracting labels.

\bibliographystyle{aasjournal}
\bibliography{monnier_gpi}

\listofchanges

\end{document}